\def\be{\begin{equation}}
\def\ee{\end{equation}}
\def\bea{\begin{eqnarray}}
\def\eea{\end{eqnarray}}
\documentclass[12pt]{iopart}
\usepackage{epsfig,graphicx}
\begin{document}

\title[Squeezed correlations of strange particle-antiparticles]{Squeezed correlations of strange particle-antiparticles}

\author{Sandra S Padula$^1$, Danuce M Dudek$^1$ and O Socolowski Jr$^2$}

\address{$^1$ Inst. F\'\i sica Te\'orica - UNESP, C. P. 70532-2, 01156-970 S\~ao Paulo, SP, Brazil}
\address{$^2$ Departamento de F\'\i sica, FURG, C. P. 474, 96201-900 Rio Grande, RS, Brazil}
\ead{padula@ift.unesp.br}
\begin{abstract}
Squeezed correlations of hadron-antihadron pairs are predicted to appear if their masses 
are modified in the hot and dense medium formed in high energy heavy ion collisions. If discovered experimentally, they would be an unequivocal evidence of in-medium mass shift found by means of hadronic probes. 
We discuss a method proposed to search for this novel type of correlation, illustrating it by means of $D_s$-mesons with in-medium shifted masses. These particles are expected to be more easily detected and identified in future upgrades at RHIC.

\end{abstract}


\section{Introduction: The hadronic squeezed states}

About ten years ago, M. Asakawa, T. Cs\"org\H{o} and M. Gyulassy\cite{acg99} finalized a model description 
for the effects of in-medium hadronic mass modification leading to correlations of boson-antiboson pairs, also known 
as Back-to-Back Correlations (BBC). This type of correlation between a particle and its own antiparticle was first noted 
by R. Weiner et al.\cite{weiner}. Within a short period after the final proposition in Ref.\cite{acg99}, 
P. K. Panda et al.\cite{pchkp01} showed that similar correlations between a fermion and a antifermion should appear, 
if their masses  were shifted in the hot and dense media formed in high energy heavy ion collisions.
 
 In both the bosonic and the fermionic cases, the in-medium quasi-particles produced in those collisions are related to the asymptotic, 
 observed particles, by means of a Bogoliubov-Valatin (BV) transformation. This transformation links the creation and annihilation 
 operators in both environments, i.e., the asymptotic operators $a$ and $a^\dagger$, to their in-medium counterparts, $b$ and $b^\dagger$. 
 The corresponding Hamiltonians are given by $H_0$ and $H_m=H_0+H'$, where $H'$ contains the parameter 
 expressing the mass-shift. 
  
 The BV transformation relating the operators $a$ ($a^\dagger$) to $b$ ($b^\dagger$) are given by
$
 a_k=c_k b_k + s^*_{-k} b^\dagger_{-k} \; ; \; a^\dagger_k=c^*_k b^\dagger_k + s_{-k} b_{-k}
$, 
where $c_k=\cosh(f_k)$ and $s_k=\sinh(f_k)$. The argument is the  {\sl squeezing parameter}, named in this way because the BV transformation is equivalent to a squeezing operation. It is written as    
$f_k=\frac{1}{2}\log\left(\frac{\omega_k} {\Omega_k}\right)$, with $\omega_k^2={\mathbf k}^2 + m^2$, $m$ 
being the asymptotic mass, 
$\Omega_k^2={\mathbf k}^2 + m_*^2$, $m_*$ being the in-medium modified mass, and $\mathbf k$ is the momentum. 
A constant mass-shift is considered here, homogeneously distributed over all the system, and related to the asymptotic mass by $m_* =m \pm \delta M$. More generally, however, it could be a function of the coordinates inside the system and the momenta, $\delta M=\delta M(|{\mathbf r}|,|{\mathbf k}|$). 
Both the bosonic and the fermionic squeezed correlations are positive, have unlimited intensity, and are also described by
similar formalisms \cite{acg99,pchkp01}. This is illustrated in Ref. \cite{pchkp01} for the case of $\phi \phi$ and 
$\bar{p} p$ pairs, evidencing the resemblance of the correlations for bosons and for fermions of similar asymptotic masses. In 
the remainder of this paper, we will focus on the bosonic case. 
 
 The effects of the shifted mass on the squeezed particle-antiparticle correlations can be understood by analyzing the
  the joint probability for observing two particles, 
$  
N_2(\mathbf k_1,\mathbf k_2) = 
   \omega_{\mathbf k_1} \omega_{\mathbf k_2} 
  \Bigl[\langle
a^\dagger_{\mathbf k_1} a_{\mathbf k_1}\rangle \langle a^\dagger_{\mathbf k_2}
a_{\mathbf k_2} \rangle + \langle a^\dagger_{\mathbf k_1} a_{\mathbf k_2}\rangle
\langle a^\dagger_{\mathbf k_2} a_{\mathbf k_1} \rangle + \langle
a^\dagger_{\mathbf k_1} a^\dagger_{\mathbf k_2} \rangle \langle a_{\mathbf k_2}
a_{\mathbf k_1} \rangle\Bigr] $, 
which results from the application of a generalization of Wick's theorem for locally
equilibrated systems\cite{gykw,sm2}; $\langle ... \rangle $ means thermal averages. 
In this expression, 
all three contributions could survive when the boson is its own antiparticle, as 
is the case of $\phi$-mesons or $\pi^0$'s. The last term  
is in general identically zero. However, if the 
particle's mass is shifted in-medium, it can contribute significantly. It is identified with the square modulus 
of the squeezed amplitude, 
{\small $
G_s({\mathbf k_1},{\mathbf k_2}) = \sqrt{\omega_{\mathbf k_1} \omega_{\mathbf k_2} } \; \langle
a_{\mathbf k_1} a_{\mathbf k_2} \rangle$. 
The first term corresponds to the product of the spectra of the two identical bosons, 
$N_1(\mathbf k_i)\!=\!\omega_{\mathbf k_i} \frac{d^3N}{d\mathbf k_i} \!=\!
\omega_{\mathbf k_i}\,
\langle a^\dagger_{\mathbf k_i} a_{\mathbf k_i} \rangle $}, 
and the second term, to the identical particle contribution, identified with the 
square modulus of the chaotic amplitude,
$ G_c({\mathbf k_1},{\mathbf k_2}) = \sqrt{\omega_{\mathbf k_1} \omega_{\mathbf k_2}} \; \langle
a^\dagger_{\mathbf k_1} a_{\mathbf k_2} \rangle$. 

The full two-particle correlation function for  $\phi \phi$ or $\pi^0 \pi^0$ can be written as 
 \be C_2({\mathbf k_1},{\mathbf k_2}) =1 + 
\frac{|G_c({\mathbf k_1},{\mathbf k_2})|^2}{G_c({\mathbf k_1},{\mathbf k_1})
G_c({\mathbf k_2},{\mathbf k_2})} + \frac{|G_s({\mathbf k_1},{\mathbf k_2})
|^2}{G_c({\mathbf k_1},{\mathbf k_1}) G_c({\mathbf k_2},{\mathbf k_2}) }. \label{fullcorr}
\ee 
In case of charged mesons, such as $D_s^\pm$ , the terms in Eq.(\ref{fullcorr}) 
would act independently, i.e., either the first and the second terms together would contribute to the HBT effect 
($D_s^\pm D_s^\pm$), or the first and the last terms, to the BBC effect ($D_s^+ D_s^-$).

In Refs.\cite{acg99,pchkp01}, an infinite system was considered. Later, a finite expanding system 
was studied, within a non-relativistic approach, assuming flow-independent squeezing parameter, which allowed for obtaining analytical solutions 
to the problem\cite{phkpc05}. Considering a hydrodynamical ensemble\cite{acg99,sm1,phkpc05} the squeezed amplitude results in 
{\small
\bea  
G_s(\mathbf{k}_1,\mathbf{k}_2)  &=&  \frac{E_{_{1,2}}}{(2\pi)^\frac{3}{2}} |c_{_0}| |s_{_0}| \Bigl\{ R^3
 \exp \Bigl[-\frac{R^2}{2}(\mathbf{k}_1+\mathbf{k}_2)^2   \Bigr] 
 +  2 n^*_0  R_*^3 
  \exp\Bigl[-\frac{(\mathbf{k}_1-\mathbf{k}_2)^2}{8m_* T}\Bigr]   
 \nonumber \\ && 
\times \exp \Big[\Bigl(- \frac{im\langle u\rangle R}{2m_* T_*} -
 \frac{1}{8 m_* T_*}  -  \frac{R_*^2}{2} \Bigr) (\mathbf{k_1} + \mathbf{k_2})^2\Big] \Bigl\},  
\label{squeezampl}\eea }  
and the spectrum in
\be
G_c(\mathbf{k}_i,\mathbf{k}_i)  =  \frac{E_{i,i}}{(2\pi)^\frac{3}{2}}  \Bigl\{|s_{_0}|^2R^3  +  n^*_0 R_*^3 
( |c_{_0}|^2 +  |s_{_0}|^2)  \exp\Bigl(-\frac{\mathbf{k}_i^2}{2m_* T_*}\Bigr)  \Bigr\}, 
\label{spec}\ee 
where {\small $R_*=R\sqrt{T/T_*}$ and $T_*=T+\frac{m^2\langle u\rangle^2}{m_*}$} \cite{phkpc05}. For the sake of simplicity, the 
system was supposed to be Gaussian in shape, with a cross-sectional area of radius $R$, and $T$ is the freeze-out temperature; $R_*$ and $T_*$ are, respectively, the flow-modified radius and temperature. The flow velocity, introduced before estimating the results in Eqs. (\ref{squeezampl}) and (\ref{spec}), was 
written as $\mathbf{v} = \langle u \rangle \mathbf{r}/R$, where the values $\langle u \rangle=0, 0.23$ or $0.5$ were used in the present work. For finite particle emissions, we considered a Lorentzian distribution, {\small $F(\Delta t)=[1+(\omega_1+\omega_2)^2\Delta t ^2]^{-1}$}, multiplying the third term in Eq. (\ref{fullcorr}). We adopt here $\hbar=c=1$. The results in Eq.(\ref{squeezampl}) and (\ref{spec}) are then introduced, respectively, in the third and first terms  of Eq.(\ref{fullcorr}), leading to the squeezed correlation function. 

\section{Illustrative results}

We previously applied the analytical results shown above to $\phi$-meson squeezed correlations, and later, to $K^+K^-$ pairs. We investigated the behavior of the squeezed correlation function for precise back-to-back pairs, i.e., for particle-antiparticle pairs emitted with exactly opposite momenta, studying $C_s(\vec{k},-\vec{k}, m_*)$ as a function of $|\vec{k}|$ and $m_*$. Preliminary results for those particles where shown in previous meetings\cite{phkpc05}-\cite{psdismd08}. 
Recent results considering $\phi \phi$ pairs from simulation are in Ref. \cite{letterphis}. 

\begin{figure}[h]
\protect
\begin{center}
\leavevmode
 \includegraphics[height=.28\textheight]{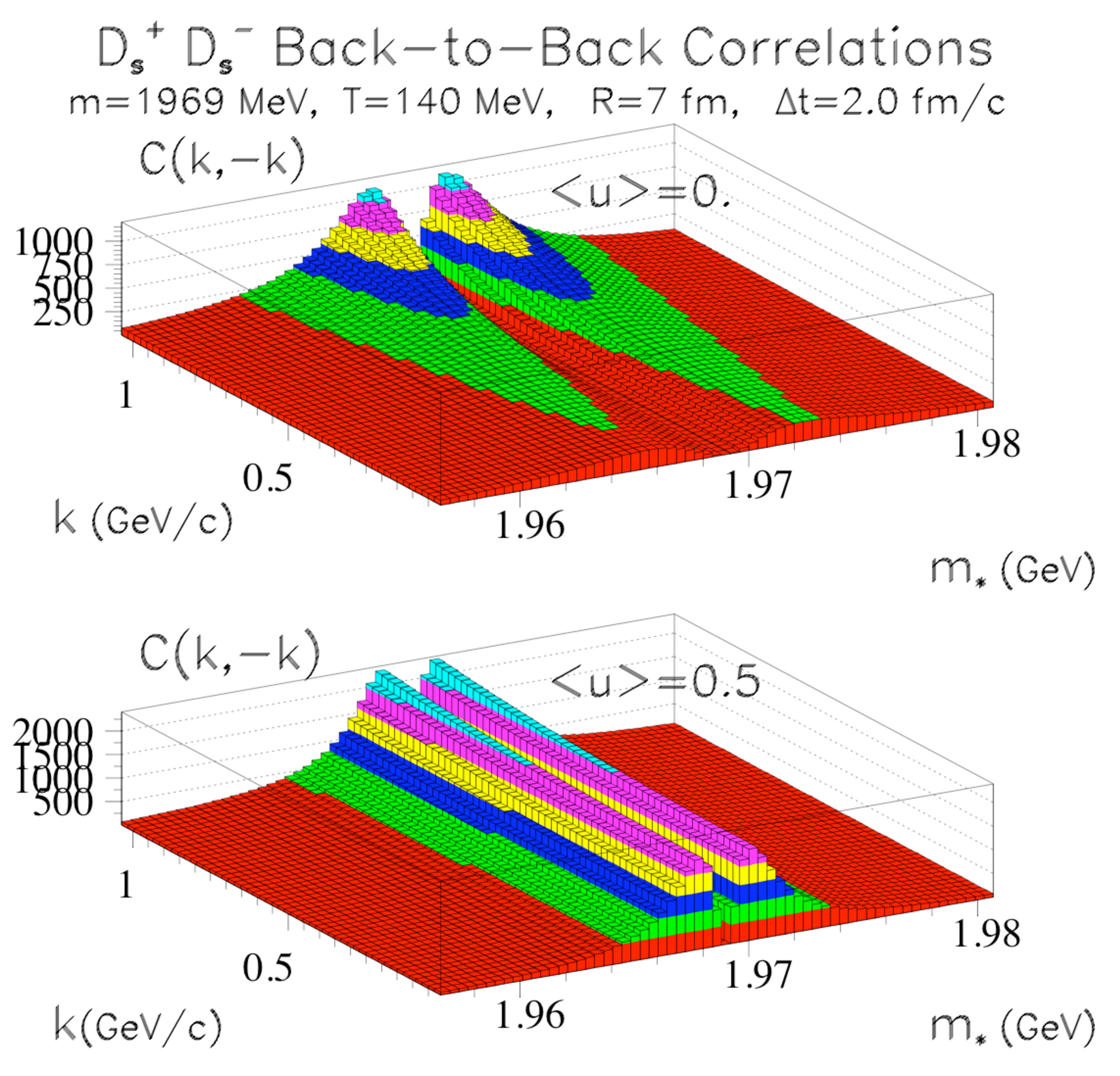}
 \includegraphics[height=.28\textheight]{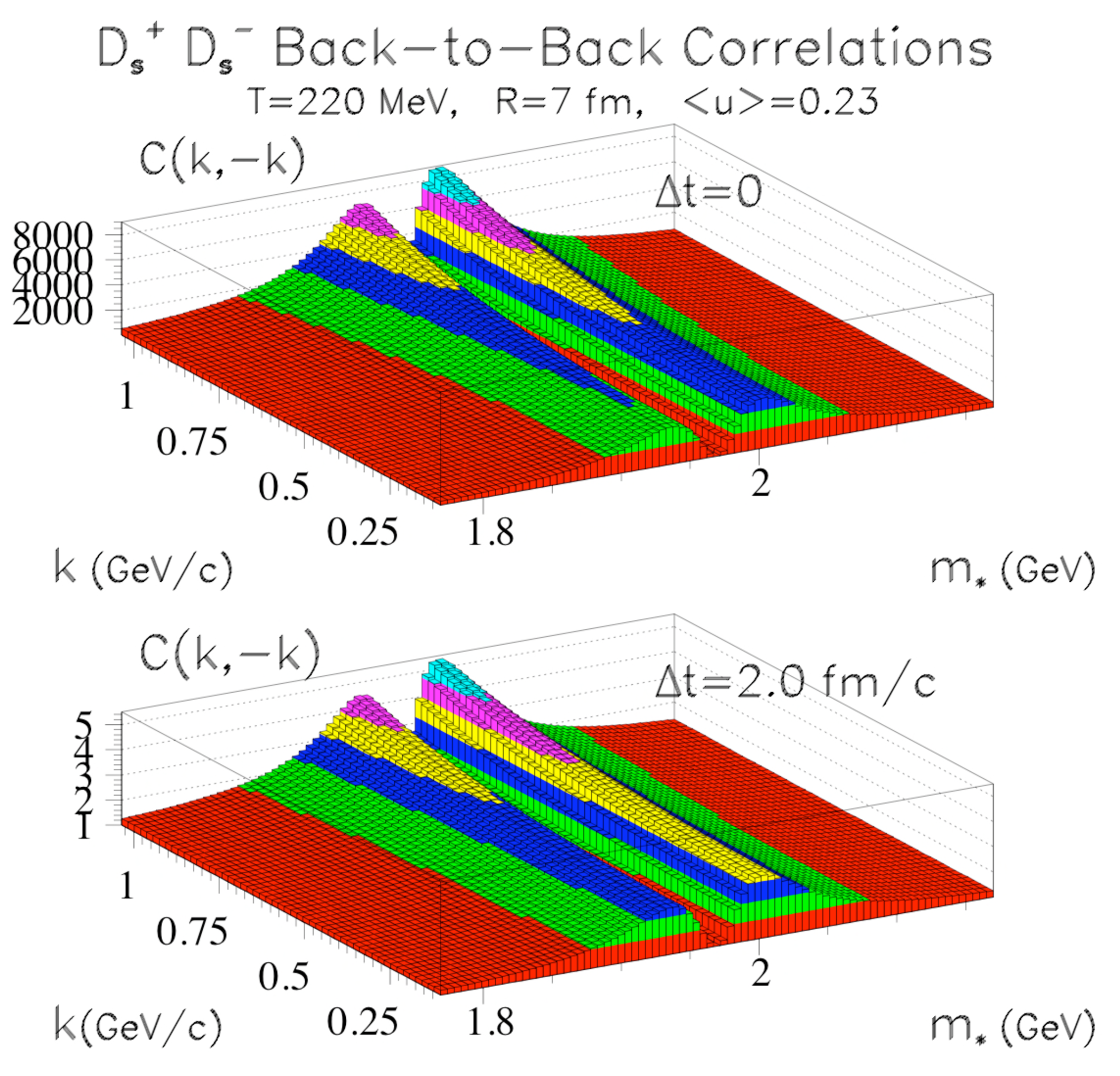}
\end{center}
\caption{{\bf Squeezed correlation function versus the shifted mass and the momenta of the particles for back-to-back $D_s^+ D_s^-$ pairs.}}
\label{cs-mstar-k}
\end{figure}

The analytical results can be applied to any other particles subjected to in-medium mass-shift and compatible 
with the non-relativistic limit considered in the formulation. Recently, STAR reconstructed $D^0 + \bar{D^0}$'s  through their decay into $K^\mp \pi^\pm$, by measuring the invariant mass distribution of those decay products\cite{STAR}. The identification of $D$-mesons could be improved after the detector's upgrade. Motivated by this possibility, and considering that charged mesons may be easier to observe, we apply that formulation to the case of $D_s^+ D_s^-$ pairs. Similar to what was previously done for $\phi \phi$ and $K^+ K^-$, we analyze the behavior of $C_s(\vec{k},-\vec{k}, m_*)$ for $D_s^+ D_s^-$ pairs in the $(|\vec{k}|, m_*)$-plane. 
 Fig. \ref{cs-mstar-k} shows that the intensity of $C_s({\mathbf k}, -{\mathbf k}, m_*)$ vs. $|{\mathbf k}|$ vs. $m_*$   
increases for decreasing freeze-out temperatures, which can be seen by comparing the two left plots with $T=220$ MeV, 
with the bottom right plot with $T=140$ MeV, in all of which $\Delta t = 2$ fm/c.  
Fig. \ref{cs-mstar-k} also illustrates another striking feature, i.e., finite emission intervals can dramatically reduce the strength of the squeezed correlation function. This can be seen by comparing the two plots in the right panel: 
a Lorentzian emission distribution with $\Delta t = 2$ fm/c has the effect of reducing the signal by about three orders of magnitude, as compared to the instantaneous emission ($\Delta t = 0$). Finally, the left panel shows how the behavior of $Cs({\mathbf k}, -{\mathbf k}, m_*)$ is affected by 
the presence of flow. We see that the growth of the signal's intensity for increasing values of $|\mathbf{k}|$ is faster in the static case than when $<u> \neq 0$.  Nevertheless, flow seems to enhance the overall intensity of $Cs({\mathbf k}, -{\mathbf k}, m_*)$ in the whole region of $|\mathbf{k}|$ investigated. Naturally, at $m_*=m_{D_s}\sim 1969$ MeV, the squeezing disappears, i.e., 
$C_s({\mathbf k}, -{\mathbf k}, m_*) \equiv 1$. 

\begin{figure}[h]
\protect
\begin{center}
\leavevmode
 \includegraphics[height=.28\textheight]{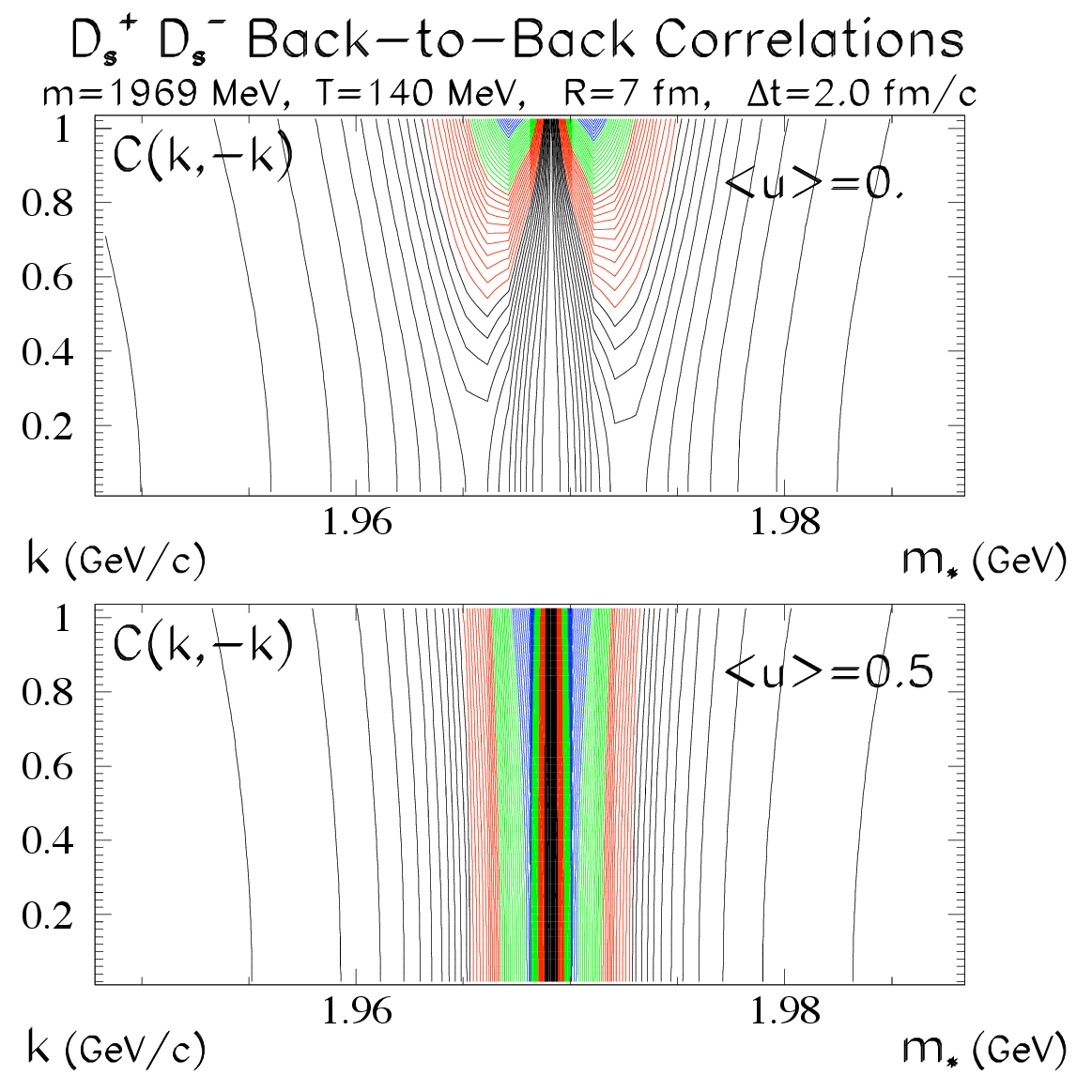}
 \includegraphics[height=.28\textheight]{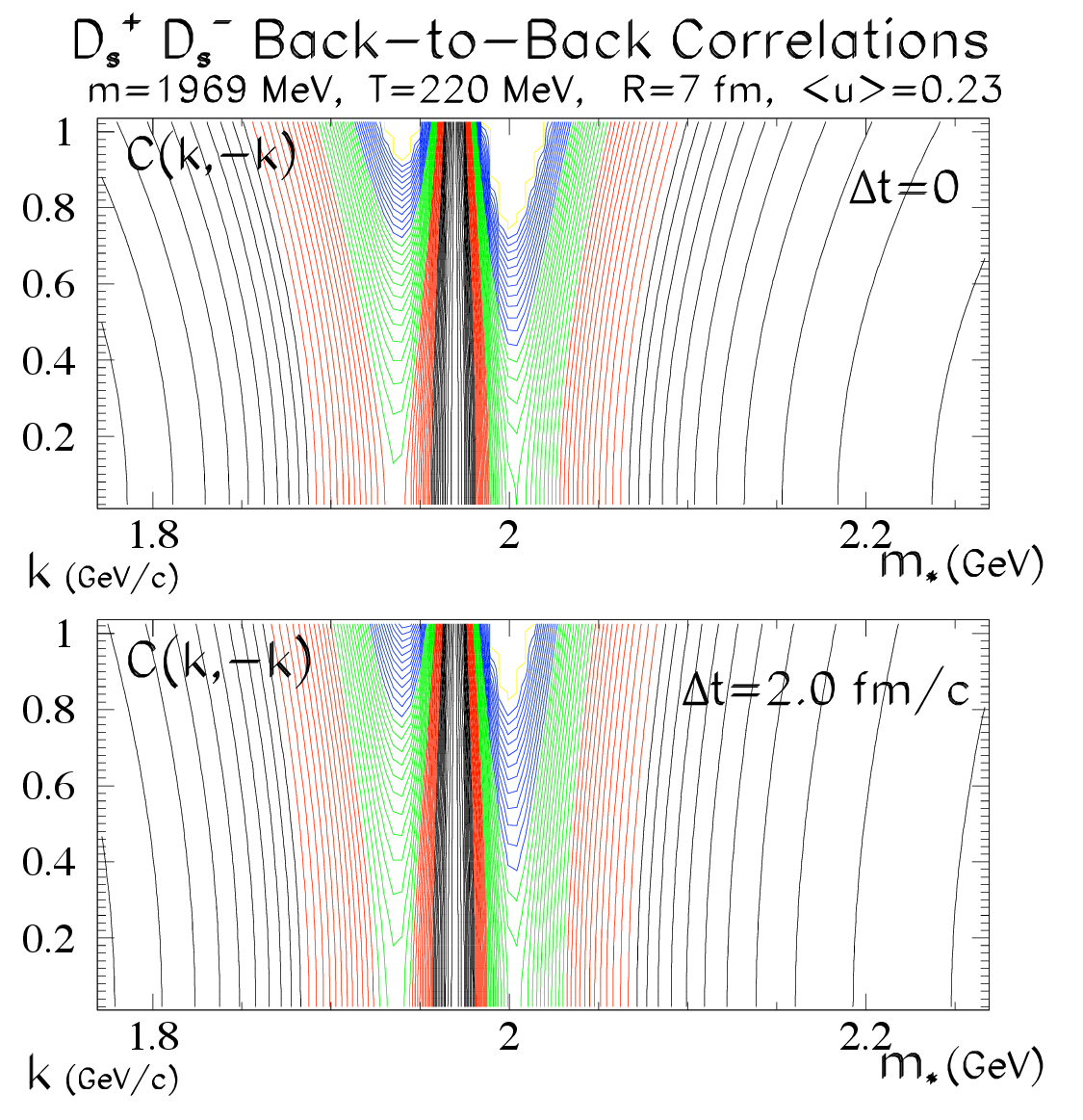}
\end{center}
\caption{{\bf Contour plots of the results of Fig. \ref{cs-mstar-k}, corresponding to the squeezed correlation functions vs. $|\mathbf{k}|$ and  $m_*$.}}
\label{cs-mstar-k-contour}
\end{figure}

A panoramic view of the variation of  $Cs({\mathbf k}, -{\mathbf k}, m_*)$ 
in the $(|\mathbf{k}|,m_*)$-plane can be better appreciated by contour plots of the previous results, as shown in Fig. \ref{cs-mstar-k-contour}. 

We can see that the location of the maxima of $C_s({\mathbf k}, -{\mathbf k}, m_*)$ for $D_s^+ D_s^-$ pairs  is 
shifted about $\approx 2 - 2.5 \%$ from the value of the particle's asymptotic mass, either above or below it. 
A similar behavior could be observed in the case of $\phi \phi$ pairs \cite{qm05}-\cite{letterphis}. However, for $K^+ K^-$ pairs, the maximum was shifted about 
$30 \%$ from the $K$'s asymptotic mass \cite{wpcf08,psdismd08}, perhaps signaling to the limit of applicability of the non-relativistic approximations considered in the model, in this case. 

The outcome properties shown above are important for understanding the expected behavior of the squeezed correlation function, within the approximations of the proposed model. However, for practical purposes of searching for it experimentally, the approach analyzed so far is not very helpful, since the modified mass of particles is not a measurable quantity, existing only inside the hot and dense medium. Besides, the measurement of particle-antiparticle pairs with exactly back-to-back momenta is unrealistic. Thus, a promising form to empirically search for the hadronic squeezed correlation was proposed, 
in analogy with HBT \cite{qm08}-\cite{letterphis},  
i.e., to measure the squeezed correlation function in terms of the momenta of the particles combined as ${\mathbf K}_{12}=\!\frac{1}{2}( {\mathbf k_1}+{\mathbf k_2})$, and ${\mathbf q_{12}}=( {\mathbf k_1}-{\mathbf k_2})$. 
In a relativistic treatment, as proposed by M. Nagy \cite{qm08}, we should construct the momentum variable as $Q^\mu_{back}=(\omega_1-\omega_2,\mathbf k_1 + \mathbf k_2)=(q^0,2\mathbf K)$. In fact, it is preferable to redefine this variable  as $Q^2_{bbc} = -(Q_{back})^2=4(\omega_1\omega_2-K^\mu K_\mu )$, whose  non-relativistic limit is $Q^2_{bbc} \rightarrow (2 {\mathbf K_{12}})^2$, correctly recovering that variable. 

\begin{figure}[h]
\protect
\begin{center}
\leavevmode
\includegraphics[height=.28\textheight]{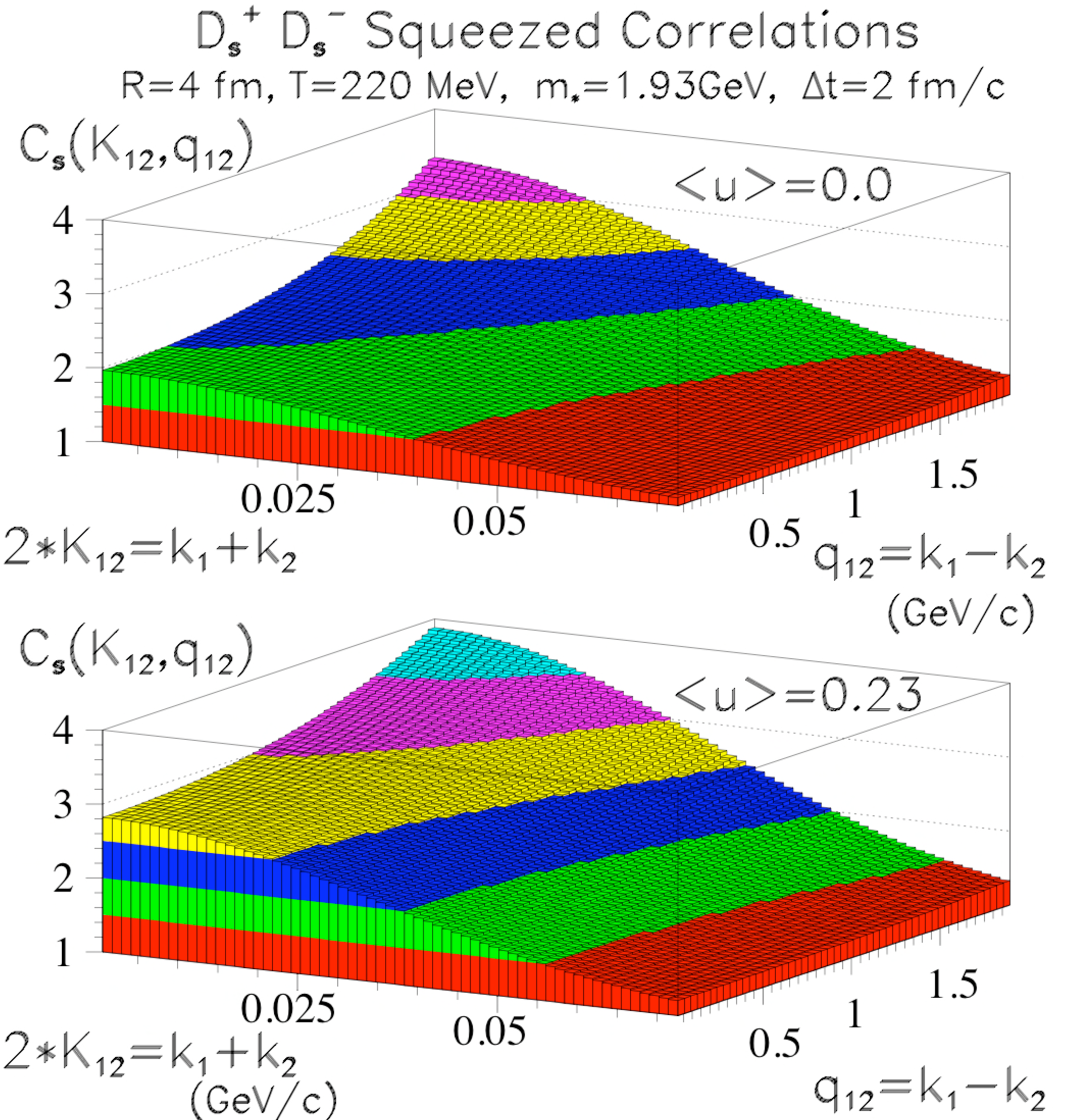}
\includegraphics[height=.28\textheight]{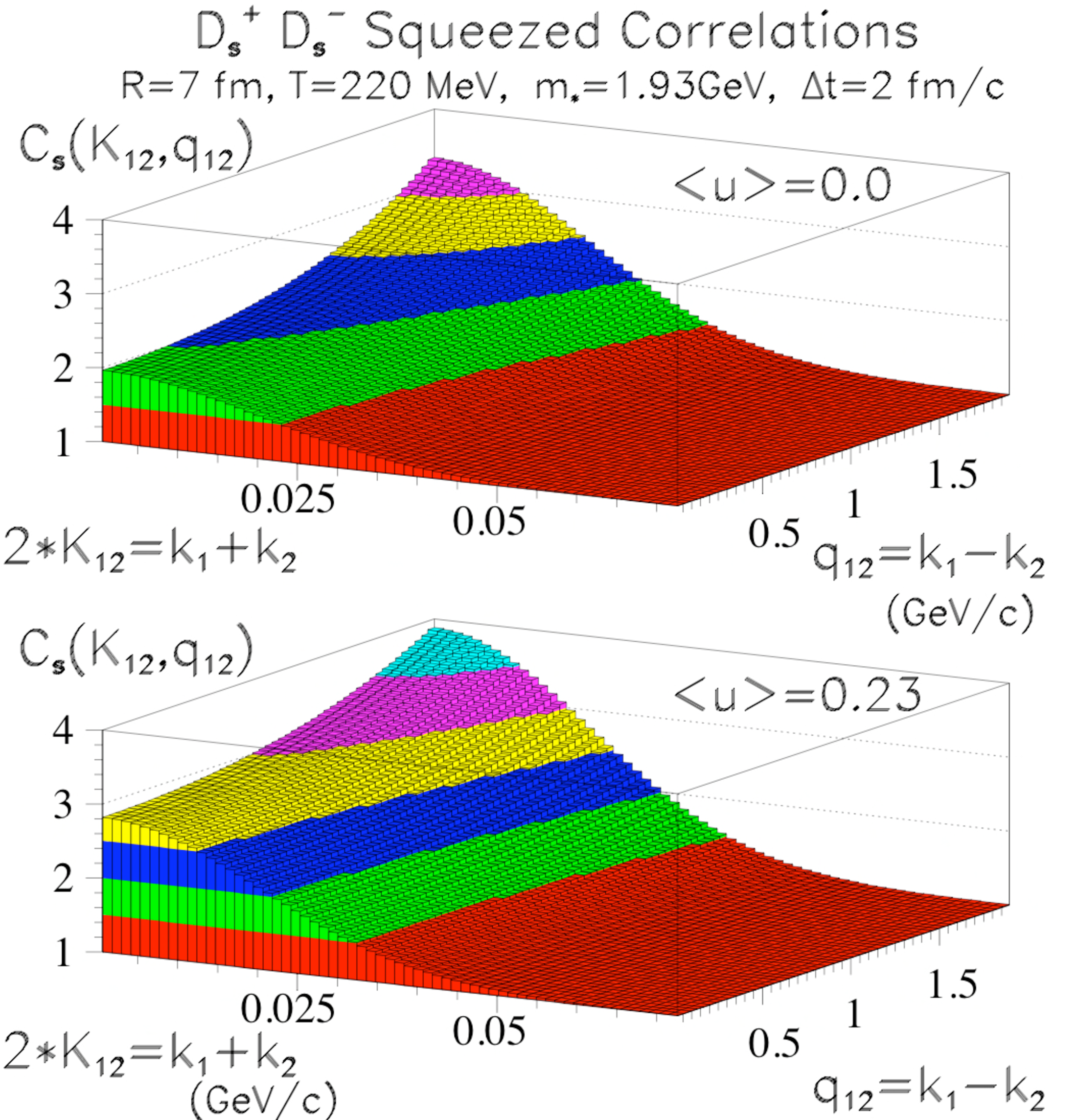}
\end{center}
\caption{{\bf Squeezed correlation functions for  $R=4$ fm (left panel) and $R=7$ fm (right panel), considering a reduction of the in-medium mass to $m_* = 1930$ MeV.}}
\label{cs-k12-q12}
\end{figure}

 The results for $C_s({2 {\mathbf K_{12}}},{\mathbf q_{12}})$ are shown in Fig. \ref{cs-k12-q12}. In the plots on the left panel, the radius of the system was fixed to be $R=4$ fm,  and on the right, to $R=7$ fm. In both, it was assumed that the mass was shifted by the amount corresponding to the lower maximum in Figs. \ref{cs-mstar-k} and \ref{cs-mstar-k-contour}, a relative reduction in the mass of about $2\%$.
 
From Fig. \ref{cs-k12-q12} we can see that radial flow ($\langle u\rangle=0.23$) does have an effect on $C_s(2 {\mathbf K_{12}},{\mathbf q_{12}})$, making it more intense if compared to $\langle u\rangle=0$, in all the investigated region of the $({\mathbf 2 K_{12}},{\mathbf q_{12}})$-plane, from about $40\%$, at low $|\mathbf q_{12}|$, to roughly $15\%$, at high $|\mathbf q_{12}|$. Fig. \ref{cs-k12-q12} also shows that the inverse width of $C_s(2 {\mathbf K_{12}},{\mathbf q_{12}})$ reflects the size of the squeezing region, being narrower for larger systems ($R=7$ fm)  than for smaller ones ($R=4$ fm). We also note that the intensity of the squeezed correlation function is high enough for stimulating its experimental search, even after applying the time reduction factor corresponding to an emission during a finite interval of $\Delta t=2$ fm/c. 

\section{Summary and conclusions}

The results  of Fig. \ref{cs-mstar-k} and \ref{cs-mstar-k-contour} showed that the squeezed correlation function survives both finite system sizes and flow, with measurable intensity. The finite emission process has a strong reduction effect in its strength, even if it happens in a sudden manner, as considered.
The plots in the right panel of Fig. \ref{cs-mstar-k}
show that the time multiplicative factor reduces $Cs({\mathbf k}, -{\mathbf k}, m_*)$ about three orders of magnitude, for $D_s^+ D_s^-$ pairs. The left panel in Fig. \ref{cs-mstar-k} shows that, if the system is subjected to flow, this could enhance the squeezed correlation signal, facilitating its potential discovery in experiments. 
The freeze-out temperature is an essential ingredient, the lower the temperature, the higher the intensity of $Cs({\mathbf k}, -{\mathbf k}, m_*)$. The expectations are that $D_s$'s would decouple early. From Fig. \ref{cs-mstar-k} we can see that, even in a high temperature limit and subjected to the time-reducing factor, 
the squeezed correlation function still has measurable intensity.  

We should remember that, if the shift in the mass is away from the value considered in the calculation leading to Fig. \ref{cs-k12-q12}, $C_s(2 {\mathbf K_{12}},{\mathbf q_{12}})$ would attain smaller values than the ones shown, but the signal could still be high enough to be searched for. 
Another important point to emphasize is that it is crucial to accumulate high statistics and look for the effect in the $3-D$ configuration shown in the plots, since projecting it into the ${\mathbf K_{12}}$-axis can drastically reduce the signal, making it more difficult to discover the hadronic squeezing effect. Finally, comparing with previous results for $\phi \phi$ \cite{qm05}-\cite{letterphis} and $K^+ K^- $\cite{qm08,wpcf08,psdismd08}-\cite{danuce-M,danuce}, we can conclude that, within this 
model, the strength of squeezed correlation function seems to grow with the asymptotic mass of the particles involved, making it even more promising to look for BBC's for heavier particles. 

\ack
We are grateful to FAPESP and CAPES for the support to participate in the SQM '09. OSJ acknowledges funding from CNPQ.

\end{document}